# Matrix multiplication using quantum-dot cellular automata to implement conventional microelectronics

**Joshua D Wood[1,3] and P Douglas Tougaw[2]**

[1] *Student Member, IEEE,* Department of Electrical and Computer Engineering, Beckman Institute for Advanced Science and Technology, University of Illinois, Urbana-Champaign, IL 61801, USA

[2] *Member, IEEE,* Department of Electrical and Computer Engineering, Valparaiso University, Valparaiso, IN 46383, USA

[3] To whom correspondence should be addressed, e-mail: jwood2@illinois.edu

**Abstract**
Quantum-dot cellular automata (QCA) shows promise as a post silicon CMOS, low power computational technology. Nevertheless, to generalize QCA for next-generation digital devices, the ability to implement conventional programmable circuits based on NOR, AND, and OR gates is necessary. To this end, we devise a new QCA structure, the QCA matrix multiplier (MM), employing the standard Coulomb blocked, five quantum dot (QD) QCA cell and quasi-adiabatic switching for sequential data latching in the QCA cells. Our structure can multiply two N x M matrices, using one input and one bidirectional input/output data line. The calculation is highly parallelizable, and it is possible to achieve reduced calculation time in exchange for increasing numbers of parallel matrix multiplier units. We show convergent, *ab initio* simulation results using the Intercellular Hartree Approximation for one, three, and nine matrix multiplier units. The structure can generally implement any programmable logic array (PLA) or any matrix multiplication based operation.

# 1. Introduction

As silicon CMOS scaling goes past the 32 nm node, short channel and quantum effects begin to degrade transistor operation substantially [1]. Moreover, further transistor scaling will require novel advances in lithographic processing and power dissipation. Therefore, quantum-dot cellular automata (QCA) is a nanotechnology solution [2] that can mitigate these concerns and extend Moore's Law predictions. In its original formulation, QCA employs the electron ground state energy in an array of zero-dimensional quantum dots (QDs) arranged in cells consisting of five QDs each at the center and the four corners of a square. When two electrons occupy each cell, Coulombic interactions and quantum mechanical tunneling cause their behavior to be highly bistable, and they tend to align in one of two diagonal arrangements. These two arrangements can be used to encode a binary "1" and a binary "0," and geometric arrangements of cells can perform useful binary operations [2, 3]. Furthermore, it is possible to take a QCA cell and raise or lower the tunneling barriers slowly, allowing encoded information to latch or transmit to another QCA cell. This slow, low energy process is known as quasi-adiabatic switching.

By use of the ground state and electron tunneling for calculation, QCA lowers power dissipation significantly, addressing one of the primary concerns with the current generation of silicon CMOS. While fabrication and room-temperature concerns exist for QCA [1], advances have been made by scanning tunneling microscopy (STM) based manipulation of dangling bonds on the hydrogen-passivated silicon surface [4] and at the interface of GaAs/AlGaAs heterostructures [5]. Electronic QDs in compounds such as graphene also show promise [6], provided top-down lithography can be controlled.

Nevertheless, the greatest challenge in harnessing the computational power of QCA is to develop new structures that can take advantage of its unique strengths. While one can study conventional microelectronic circuits to determine what is possible, it is necessary to transfer these ideas to the new QCA architecture. Two of the fundamental systems critical to digital computation are the programmable logic array (PLA) and its counterpart, the field programmable gate array (FPGA). Both of these microelectronic devices lend themselves to matrix implementations, one of our motivations for investigating a QCA implementation of binary-based matrix multiplication.

This paper demonstrates the use of QCA devices to implement matrix multiplication, one of the most important arithmetic operations in linear algebra and quantum mechanics [7-9]. Matrix multiplication has numerous applications in the fields of physics, chemistry, and engineering, and it is the foundation upon which most of modern computer animation is built [10]. A QCA matrix multiplier could be an important element of future quantum computing systems.

# 2. Matrix multiplier model

Figure 1 shows our QCA configuration, composed of five QDs with a nearest neighbor spacing of 20 nm. There are other QCA configurations besides the five QD formulation we present here, particularly, four QD and six QD QCA [5, 11, 12]. The four QD structure appears easier to construct by virtue of one fewer QD in the QCA cell [5]. Conversely, molecular QCA using the six QD formulation shows promise for switching-based implementations [11, 13]. Regardless of these benefits, we examine the five QD structure due to its sharper bistable cell-cell response compared against the four and six QD configurations [14]. Our cell to cell separation is 60 nm, allowing for excellent bistable switching from one polarization (P = –1) to another (P = +1) by cellular interaction, shown in Figure 1(b). We note that this particular QCA spacing is achievable using modern top-down lithographic techniques, making a viable system if it were to operate at cryogenic temperatures. When we place multiple QCA cells together, we can obtain all the necessary building blocks used in modern microelectronics, given in Figure 2 for wires, inverters, and the AND gate. The AND gate of Figure 2(c) is actually a generalization of the majority gate, which will be discussed further later.

To motivate the mathematics behind our multiplier, we set the product **C** of two matrices **A** and **B** to be:

$$\mathbf{C} = \mathbf{AB} = \begin{bmatrix} \mathbf{C_1} \\ \mathbf{C_2} \\ \mathbf{C_3} \end{bmatrix} = \begin{bmatrix} a_{11}b_{11} + a_{12}b_{21} + a_{13}b_{31} & a_{11}b_{12} + a_{12}b_{22} + a_{13}b_{32} & a_{11}b_{13} + a_{12}b_{23} + a_{13}b_{33} \\ a_{21}b_{11} + a_{22}b_{21} + a_{23}b_{31} & a_{21}b_{12} + a_{22}b_{22} + a_{23}b_{32} & a_{21}b_{13} + a_{22}b_{23} + a_{23}b_{33} \\ a_{31}b_{11} + a_{32}b_{21} + a_{33}b_{31} & a_{31}b_{12} + a_{32}b_{22} + a_{33}b_{32} & a_{31}b_{13} + a_{32}b_{23} + a_{33}b_{33} \end{bmatrix} \quad (1)$$

made up of row vectors $\mathbf{C_i}$. Thus, for each row vector $\mathbf{C_i}$, a total of nine multiplications and six additions are necessary. For a binary system, the multiplications and additions for normal matrix multiplication are replaced by the two digital logic operations AND and OR, respectively, in a process known as binary matrix multiplication. This is described in more detail in the supplementary information. Although the QCA system described in this paper is designed to perform binary matrix multiplication, such a system could be used to perform matrix multiplication with integers, real numbers, or even complex numbers if it were parallelized or if the calculation were serialized. Our basis QCA matrix multiplier builds off other QCA multiplication formalism, including parallel QCA multiplication [15], pipelined array multipliers [16], and serial/parallel multiplication [17]. Furthermore, we leverage QCA structures for binary operations that were previously shown [18, 19].

Previously, it was shown that QCA devices could successfully model conventional sequential circuits through the use of quasi-adiabatic switching [20, 21]. The switching process raises and lowers standard QCA tunneling barriers to control the flow of data throughout the QCA structure. Figure 3 shows the four different quasi-adiabatic switching states: locking, locked, relaxing, and relaxed [22]. The transition from locking, locked, relaxing, and relaxed states is dependent on the magnitude of the QCA cell's tunneling barriers. Relaxing the cell allows for a cell reset, and as the cell transitions from the locking to locked state, an input is applied, changing the cell value [20]. At some points, we will leave the cells in this locked state for more than one clock cycle, which allows those cells to store their contents without regard to changes in neighboring cells. This will be referred to as a "blocking" state. These states will be used in the implementation of our QCA matrix multiplier. Thus, this latching notion employed in quasi-adiabatic switching functions similarly to a clocked multiplexer [20]. We employ zone-based clocking schemes for the quasi-adiabatic switching regions [20, 22], avoiding the spatial issues which result when the clocking regions have the same lateral dimension as the QDs themselves.

A QCA majority cell is the primary computational unit in QCA calculations. It can function as an AND gate or an OR gate, depending on its nearest neighbor cells, as shown in Figure 2(c). If one of these cells is polarized to +1, then the majority cell will be an OR gate; otherwise, if it is polarized to –1, then it will be an AND gate. Fixing the polarization for AND and OR gates makes these cells invariant to quasi-adiabatic switching. Thus, these cells are in the locked state. We will use these locked cells for the AND and OR operations necessary in binary matrix multiplication.

The QCA matrix multiplier (MM) model presented here implements binary matrix multiplication through the use of majority gates and quasi-adiabatic switching. From a conventional electronic standpoint, it uses an AND gate, an OR gate, a gated D-latch, and a tri-state buffer. Figure 4 shows this digital circuit along with its QCA implementation. We pass any data to be processed into A and B. The QCA wire A is an input only, unidirectional data line. Conversely, the QCA wire B functions as a bidirectional input/output line, depending on the state of the tri-state buffer region. Normally, this region will be a QCA blocking region, consistent with the high impedance state of a tri-state buffer. Initially, the blocking region will have to be lowered to allow for a reset of the MM memory, as detailed later. Once the reset signal propagates through the blocking region, the region's tunneling barriers are raised to begin the MM calculation. As a result, the QCA MM memory will take on the value of the reset signal. The entire reset process takes three clock cycles to perform. We note that it is also possible to add another QCA wire (which we term "R" in the supplemental material) for resetting the MM memory. This would obviate two of these three clock cycles, improving the MM operating time. Nevertheless, the additional QCA cells can be problematic when scaling the MM to larger numbers of units, causing issues with interconnect wiring and stray charges. Moreover, the added QCA wire gives the design less compactness and flexibility.

In Figure 4(b), we show a region called an OR loop. This loop is a series of three quasi-adiabatic QCA clocking regions that allow the result from a previous AND operation to be ORed with the next

AND operation. The use of three clocking regions is to synchronize the data from the last AND to the current AND operation. Within the OR loop, data will pass from region to region until it arrives back at the QCA OR gate. Therefore, this loop functions like memory for the QCA MM, lending credence to its representation as a gated D-latch.

We can pass the desired data vectors $C_i$ extracted from eq. 1 into the MM. The total number of operations required for actual matrix multiplication is based on the minimum of $m_1n_1$ and $m_2n_2$, the dimensions of the two matrices being multiplied ($m_1 x n_1$ and $m_2 x n_2$ matrices, respectively). In the case of eq. 1, the two matrices' dimensionality implies 9 total calculations and clock cycles. After these operations are completed, the blocking region of Figure 4 will relax, and the result of the calculation will pass to C. Note that here C is equivalent to the original B data line, emphasizing the bidirectionality of that QCA wire. The matrix multiplication result shows at the output after two clock cycles. Thus, the QCA matrix multiplier will have completed the multiplication of one row vector **A** and one column vector. The multiplications will continue until we calculate the entire resultant matrix **C** given in eq. 1.

### 3. Matrix multiplier simulation

Throughout the rest of the work, we assume that we are attempting to multiply two 3 x 3 matrices **A** and **B**. In our QCA model, there are five quasi-adiabatic, zone-based switching regions, named regions A through E, correspondingly, shown in Figure 4(b). A standard QCA MM device will raise and lower these regions' potentials accordingly; this promotes or decreases electron tunneling within the individual cells and latches or blocks values in those regions [20]. Therefore, this will control the sequential flow of data (i.e., state transitions) throughout the MM. Figure 5 gives the timing diagram for the QCA matrix multiplier's quasi-adiabatic switching regions, and we include the state transition table in the supplementary information. Given the dimensions described earlier, the clock signals can be implemented using metal back-gates near the plane of the cells, shown schematically in the supplemental information. By using molecular QCA and a sinusoidal clocking field transverse to the QCA cells, one could achieve a wave-based clocking scheme [11]. Nonetheless, it is not clear how a zone-based clocking scheme could be implemented if this device were reduced to a molecular implementation.

The following results are all *ab initio* numerical simulations of self-consistent QCA ground states using the Intercellular Hartree Approximation and a Hamiltonian described by second quantization operators [2]. Initially, the value of **B** will be binary "0" (P = −1) to reset the data memory in the D latch. The value of **A** here is irrelevant, but we will set it to binary "0". Following the timing diagram of Figure 5, the tri-state buffer (blocking region) will be lowered to allow the **B** data to enter the D latch region. On the next clock cycle, the "0" value will propagate into the blocking region, while the tunneling barriers in region D of the OR loop will be lowered. The subsequent clock cycle will pass the value from region E into region D, and the blocking region's tunneling barriers will be raised again. Figure 6 shows simulation results for all these reset stages. We now demonstrate the standard operating procedure for the QCA matrix multiplier using **A** and **B** matrix entries of "0" and "1", respectively. After the multiplier's memory has been successfully reset, actual data calculations for matrix multiplication can occur. The inputs of "1" and "0" are passed through region A at inputs **A** and **B**, respectively. On the next cycle, these two values are ANDed and the result ("0") is propagated through region B. This AND result is ORed with the value contained in the OR loop, which is "0" from the reset of the multiplier. That value of "0" is propagated through region C. On the next cycle, the value is propagated into the next region of the OR loop, region D. Since for a 1 x 1 matrix multiplication one AND and one OR operation are necessary, region E, the blocking region, will lower after the value is propagated into region D. The cells of region E will then drive the value out through the region A cells. The output of the operation ("0") will show up at the output **C**. Figure 7 details all of these steps.

To perform a 3 x 3 matrix multiplication with a single matrix multiplier unit, the reset will require three clock cycles, and the three multiplications and additions will require an additional 12 cycles. The output stage will require 2 cycles to lower the barrier in region E and allow tunneling to the output in region A. Therefore, for one resultant entry, 17 clock cycles will be necessary. For a 3 x 3 matrix multiplication, there will be 9 resultant entries in **C**; thus, the total cycles required for the structure is 153.

Nonetheless, this model will take very little area to actually implement and will require significantly lower amounts of interconnects amongst its timing layers.

**4. Matrix multiplier parallelization**

We now extend the one QCA multiplier unit in Figure 7 to three and nine multiplier units. A three multiplier unit model will be able to calculate a row of the resultant matrix **C**. Given the row vector $\mathbf{A}_1$ and the test matrix **B**

$$\mathbf{A}_1 = \begin{bmatrix} 1 & 1 & 1 \end{bmatrix} \quad (2)$$

$$\mathbf{B} = \begin{bmatrix} 0 & 1 & 1 \\ 1 & 0 & 1 \\ 0 & 0 & 0 \end{bmatrix} \quad (3)$$

This will give a resultant row vector $\mathbf{C_1}$, the value given in equation 1. Consequently, row vectors of $\mathbf{A}_2$ and $\mathbf{A}_3$ can give values for $\mathbf{C}_2$ and $\mathbf{C}_3$, respectively. The value of the resultant $\mathbf{C_1}$ is

$$\mathbf{C}_1 = \begin{bmatrix} 1 & 1 & 1 \end{bmatrix} \quad (4)$$

The three multiplier model follows the same stages that the one multiplier model uses. Initially, it resets the OR loops to 0 before any calculations occur. This takes three cycles; after this reset stage, the first bit of $\mathbf{A}_1$ ($a_{11}$ in eq. 1) and the entries of the first column of **B** ($b_{11}$, $b_{21}$, $b_{31}$ in eq. 1) are passed into the three multiplier device. These entries are ANDed together and the result is stored in the OR loop, a process which takes four clock cycles (see supplementary information). The process continues for the second and third bits in $\mathbf{A}_1$ ($a_{12}$, $a_{13}$ in eq. 1) and the entries of the second and third columns of **B**. The value for each entry of $\mathbf{C}_1$ in equations 1 and 4 will be located in the three OR loops. These values will be propagated to the outputs using two more clock cycles. Figure 8 shows the simulated verification of $\mathbf{C}_1$ of eq. 4.

For the three multiplier unit, 17 cycles are necessary for row vector calculation, the same as the single multiplier model. Since there are 3 row vectors within the entire resultant matrix **C**, the total number of cycles necessary for a 3 x 3 matrix multiplication is 51. This model is a middle ground between the aforementioned one multiplier model, which had the least area but took the most time, and the following nine-multiplier model, which will take the largest area but has the shortest computation time.

We now will extend our model to have nine multiplier units and verify its operation. For the given test matrices **A** and **B**

$$\mathbf{A} = \begin{bmatrix} 0 & 0 & 1 \\ 1 & 0 & 1 \\ 0 & 0 & 0 \end{bmatrix}, \mathbf{B} = \begin{bmatrix} 1 & 0 & 1 \\ 1 & 1 & 1 \\ 0 & 0 & 1 \end{bmatrix} \quad (5)$$

the resultant output matrix **C=AB** is

$$\mathbf{C} = \mathbf{AB} = \begin{bmatrix} 0 & 0 & 1 \\ 1 & 0 & 1 \\ 0 & 0 & 0 \end{bmatrix} \quad (6)$$

The system will require 17 cycles to operate: three cycles for a reset, 12 cycles for the computation of the 3x3 resultant element, and two cycles to propagate that result to the output. We give the simulated result for this model in the supplementary information. The output elements (based on the notation in equations 1, 2, and 3) correspond correctly to what is expected in eq. 6. Therefore, the entire matrix can be calculated in only 17 cycles. This model will clearly take the largest area, but the computation can be done in the shortest amount of time. Larger MM structures allow for more generality in implementing conventional microelectronic circuits, as there is no throughput necessary for serializing the input data.

We now assume that the QCA cell-to-cell size is 60 nm and the clocking zones are layered as shown in the supplemental information. From this, we can derive a first order approximation the delay times associated with the quasi-adiabatic clocking zones, which are related to the overall tradeoff between MM footprint and timing. We examine MM clocking metal layers that are made of copper ($\rho = 1.68 \times 10^{-6}$ $\Omega \cdot cm$) and separated by 2 nm of $SiO_2$ ($\varepsilon_r = 3.9$). If the area footprint is 16 x 10 QCA cells from the one multiplier in Figure 4(b), the overall clocking line parasitics are $R = 2.1\,\Omega$ and $C = 10.4$ fF, giving a delay time of $\tau \simeq 20$ fs. The delay time scales quadratically with number of MMs, with $\tau \simeq 180$ fs for 3 MMs and $\tau \simeq 1.6$ ps for 9 MMs. Thus, one must forego a larger delay when using longer interconnects in an ensemble MM structure in addition to the larger area footprint. Nevertheless, the high level of parallelism in large MM ensembles can augment the fundamental clock line delay. Additionally, conventional microelectronics are clocked at frequencies much lower than the delay limits given here, potentially making clocking interconnect cross-talk less problematic.

## 6. Implementing conventional microelectronics

To extend conventional microelectronic circuits into the QCA paradigm, we discuss the implementation of the programmable logic array (PLA) using the QCA MM. Assume we have three assertable inputs A, B, and C, which are used to implement a standard digital function. Setting a row vector **X** of these conventional inputs:

$$\mathbf{X} = \begin{bmatrix} A & B & C \end{bmatrix} \quad (7)$$

Figure 9(a) gives a PLA for the digital circuit F = AB + BC + AC. We define a node matrix **N** based on the locations of PLA connections, with "1" and "0" denoting a connection and disconnection. Each row of the PLA corresponds to a column of the node matrix **N**; therefore, the PLA matrix is the transpose of the node matrix. Breaking this node matrix **N** into its column vectors, following Figure 9(b):

$$\mathbf{N} = \begin{bmatrix} \mathbf{N}_1 & \mathbf{N}_2 & \mathbf{N}_3 \end{bmatrix} \quad (8)$$

By multiplying the negation of **X** with $\mathbf{N}_1$ from Figure 9(a) (or, generally, $\mathbf{N}_i$) and taking its negation, we get a sum of products (SOP) by DeMorgan's Theorem:

$$\overline{\left(\bar{\mathbf{X}}\mathbf{N}_1\right)} = \overline{\begin{bmatrix} \bar{A} & \bar{B} & \bar{C} \end{bmatrix} \begin{bmatrix} 1 \\ 0 \\ 1 \end{bmatrix}} = \overline{\overline{A} + \overline{C}} = AC \quad (9)$$

We can continue this procedure until we solve for the entire node matrix **N**. Thus, we modify our QCA multiplier cell to account for these negations automatically, as shown in Figure 9(c). Consequently, we can generalize this to any size PLA.

An added benefit of the MM-based PLA is that we avoid any standard PLA output logic macrocells by the generalized node matrix **N**. Thus, these PLA structures could be significantly simpler in terms of layout compared to those discussed in the literature [23]. Furthermore, previous attempts at PLA structures using QCA made use of complex, ~4 $\mu m^2$ QCA macrocells [24], leading to difficulties when realizing devices in small areas without significant QCA cell crosstalk. Current QCA representations of the field-programmable array of logic (FPGA) use lookup tables, memory loops, and configurable logic blocks [25], all of which take up substantial area and can affect latency. If one passes the input data for the QCA MM serially, the MM can have a small area footprint, allowing for faster data calculation in the PLA or FPGA structures.

## 7. Conclusions

We developed a new QCA structure, the QCA matrix multiplier. Our structure is based on QCA majority gates, data flow by quasi-adiabatic switching, an OR loop memory construct, and a tri-state buffer blocking region. With a bistable five QD cell QCA geometry, we performed an *ab initio* simulation of one, three, and nine MMs, showing how the structure is parallelizable. The MM can multiply two general N x M matrices. Further, we demonstrated the tradeoff between MM area footprint and calculation time. Our structure's ability to implement PLA structures efficiently by matrix multiplication will help transition conventional microelectronic structures into the QCA paradigm when CMOS silicon scaling reaches its limits. Moreover, our system can perform any calculation that requires matrix multiplication, such as those in computer animation or in quantum encryption. Furthermore, although this structure was designed for and simulated using quantum-dot cellular automata, the approach presented here seems to be appropriate for other similar systems.  In particular, the use of a five-dot cell does not seem to be crucial, and it seems likely that this approach would also work using magnetic QCA systems. Employing the MM with room-temperature QD operation should allow for sophisticated, programmable QCA computers to be made.

**Acknowledgment**
JDW acknowledges financial assistance from the Army Research Office through the National Defense Science and Engineering Graduate Fellowship and engaging discussions with S. W. Schmucker. JDW and PDT both thank E. W. Johnson and K. J. Olejniczak for thoughtful comments during manuscript preparation.

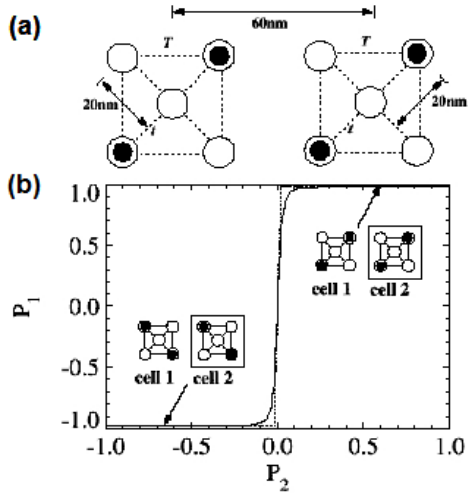

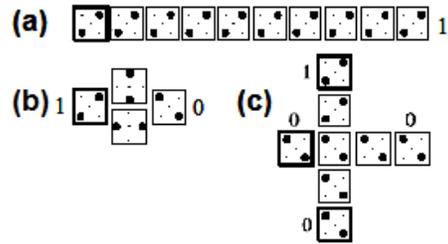

**Figure 1.** (a) Five-dot QCA cell geometry. Each cell is composed of five dots located at the center and the four corners of the square. Nearest-neighbor spacing between dots is 20 nm, and the distance between neighboring cells is 60 nm. (b) This geometry leads to a highly bistable interaction between neighboring cells, which can be utilized to encode binary '0's and '1's.

**Figure 2.** Fundamental QCA building blocks. (a) The binary wire. (b) The inverter. (c) The AND gate.

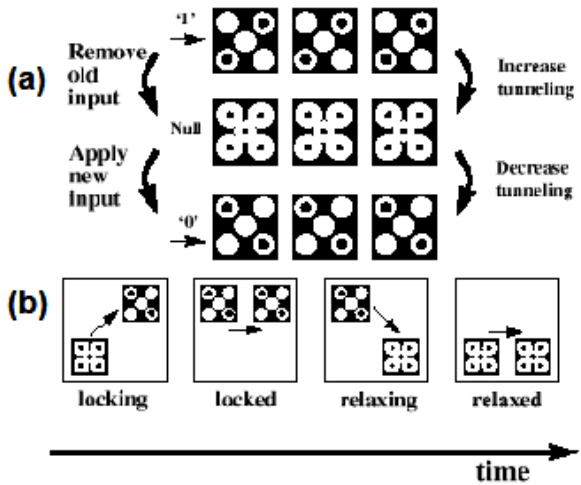

**Figure 3.** Quasi-adiabatic switching. (a) The tunneling barriers within a cell are lowered, allowing the electrons to spread more evenly over all five dots. At the same time, the input is removed. As the new input is then applied, the tunneling barriers are increased, and the cells smoothly transition to the correct state corresponding to the new input. (b) An alternative representation of the quasi-adiabatic switching process that illustrates the four distinct stages of QCA switching (locking, locked, relaxing, and relaxed).

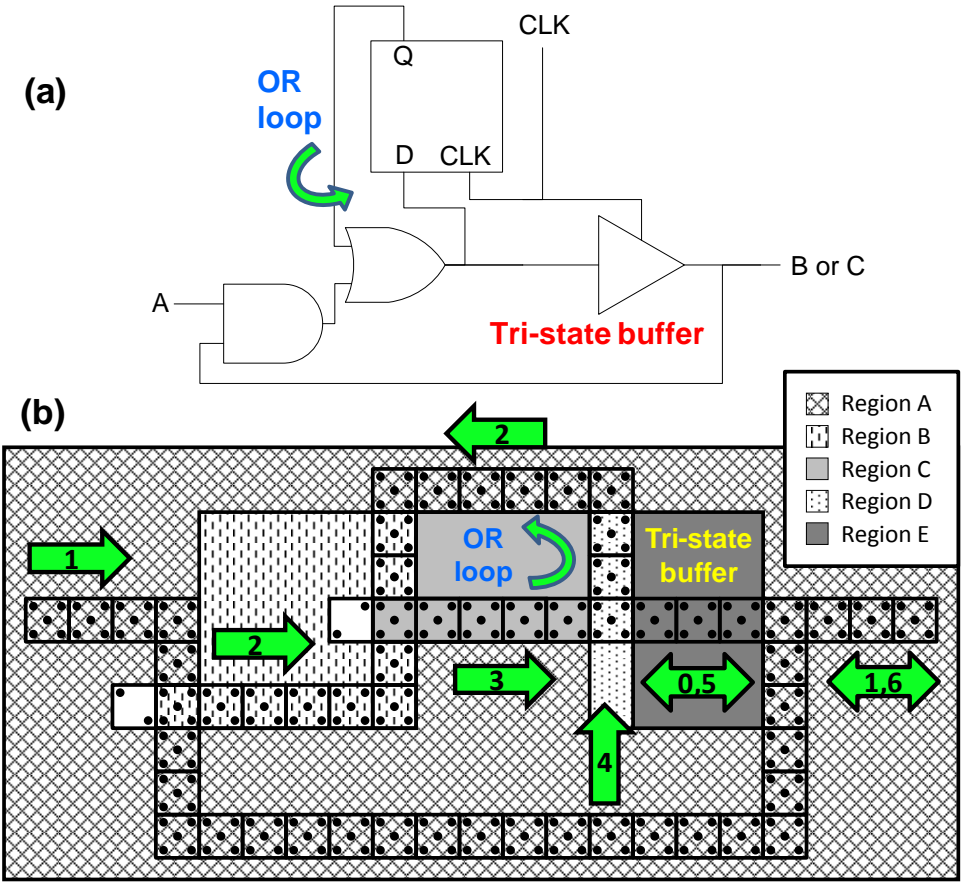

**Figure 4.** QCA matrix multiplier (MM) structures. (a) Pseudo-digital schematic of the QCA MM. Data flows from the left and right via the inputs A and B, and the output operation flows out to C. Thus, the blocking region E acts as a tri-state buffer. (b) MM configuration using a five QD cell for QCA calculations. The arrows show the data flow, with 0 corresponding to the OR loop reset, 1 to the input of the first data bit, and so forth, until the data drives out at 6. The inset shows the different clocking regions are for their respective shadings.

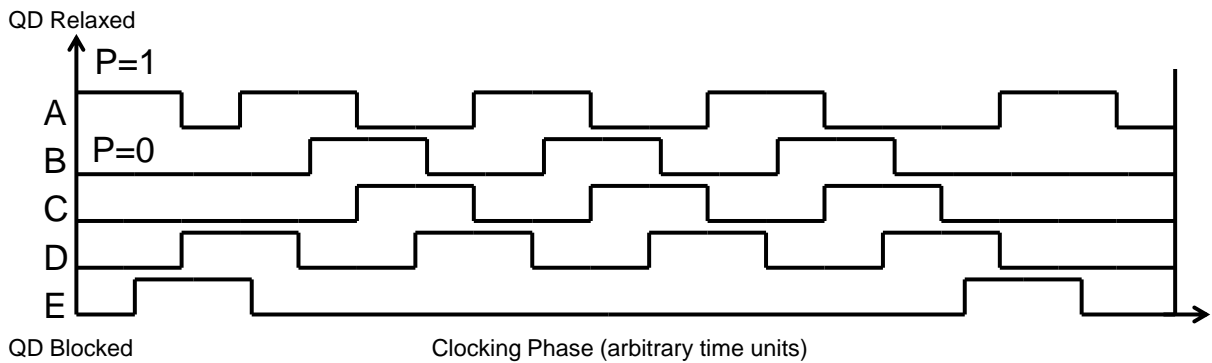

**Figure 5.** Timing diagram for the different regions in the QCA multiplier. When the signal is high, then the tunneling barriers are lowered to propagate a signal along the QCA wire (probability of P=1). Conversely, when the signal is low, the tunneling barriers are raised, blocking signal propagation (P=0). We assume there to be electron confinement within the QD, which is reasonable at low temperatures. The operation speed is dependent on the clocking rate, typically ~10ns.

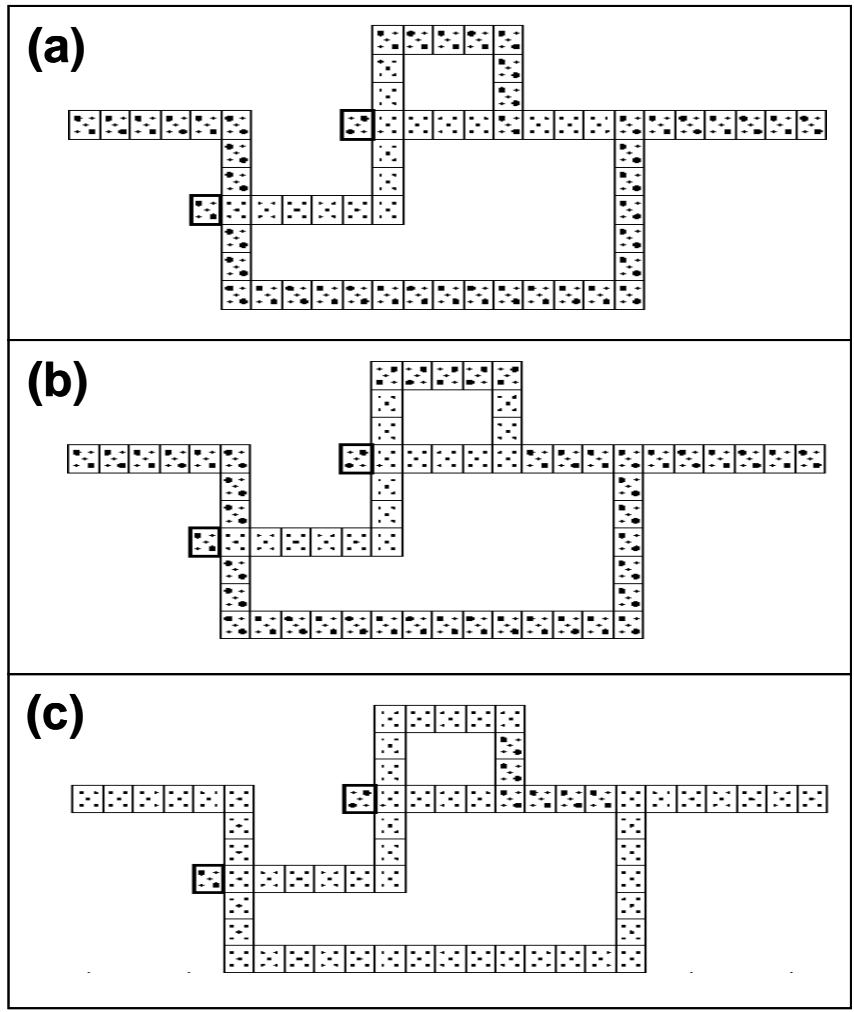

**Figure 6.** Reset stages for a single QCA MM. (a) Initial reset stage, where the inputs A and B are set to 0. The values on the cells at the top of the OR loop are due to stray charge Coulombic interaction. (b) Stage where the input value of 0 is pushed through the blocking region, region E. (c) Stage where the 0 value has successfully propagated through region E into the OR loop, region D. This will reset the OR loop value (i.e., reset the D latch).

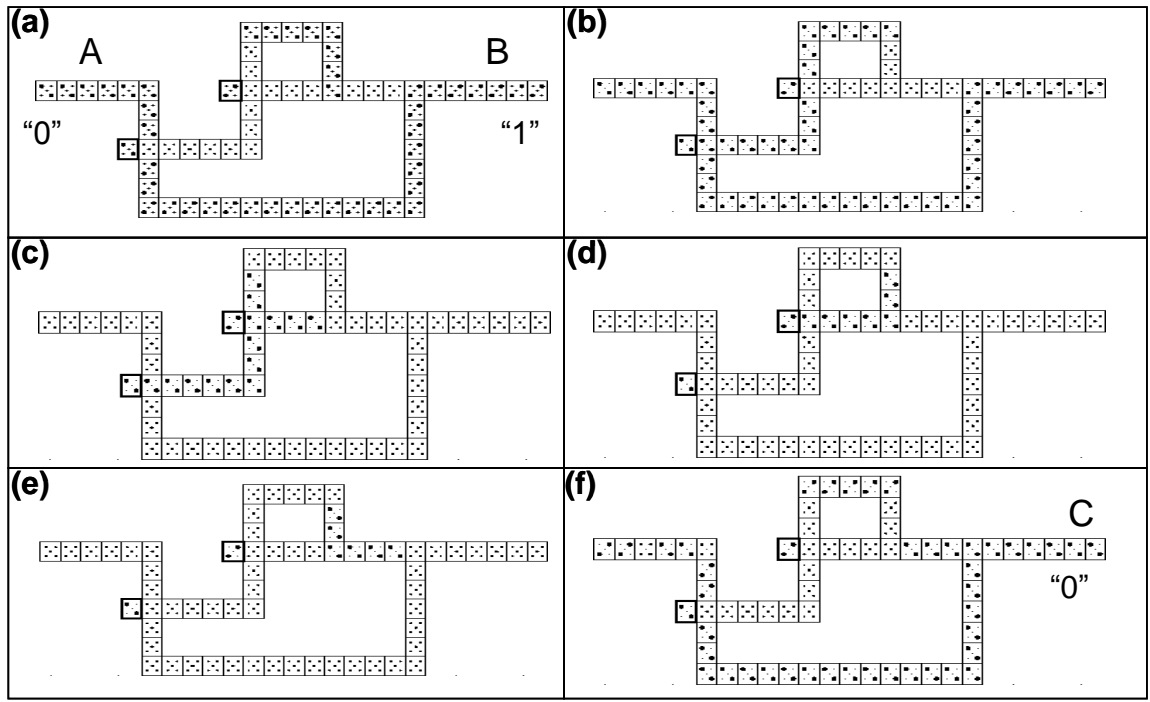

**Figure 7.** QCA matrix multiplier steps for the input bits "0" and "1" for vectors **A** and **B**, which are scalars here (1 x 1). (a) Driving a "1" bit at A and a "0" bit at B in clocking region A (see Figure 4(b) and Figure 5). (b) AND of "1" and "0" at the first majority gate, with a propagation of "0." (c) OR of the reset value of "0" with the "0", giving a "0" result. (d) Propagation of "0" value around OR loop. (e) Lowering of the pseudo-multiplexing, blocking region (region E), to propagate the output. (f) The output of the operation is shown at the output C, namely "0." The cell misalignments near the left end of the device at this point are caused by the cells in region A locking when no input is applied. This does not affect the result, but it does result in a semi-random orientation of those cells until the next input is applied.

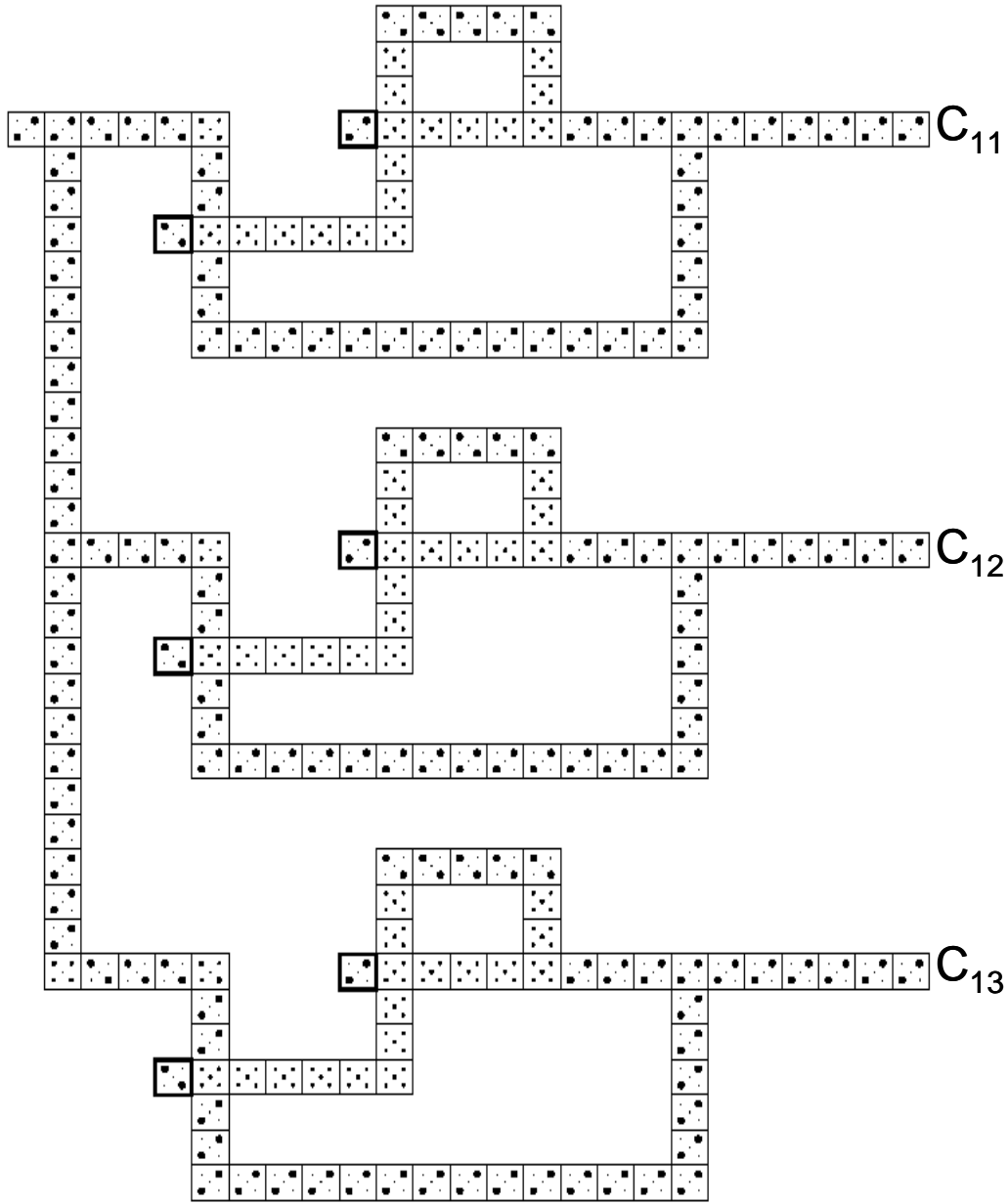

**Figure 8.** Simulated output stages of the three multiplier model. Notice that the output values are $C_{11}=1$, $C_{12}=1$, and $C_{13}=1$. Stray charges at the input **A** give the quasi-random results at those cells, even though they do not affect the output.

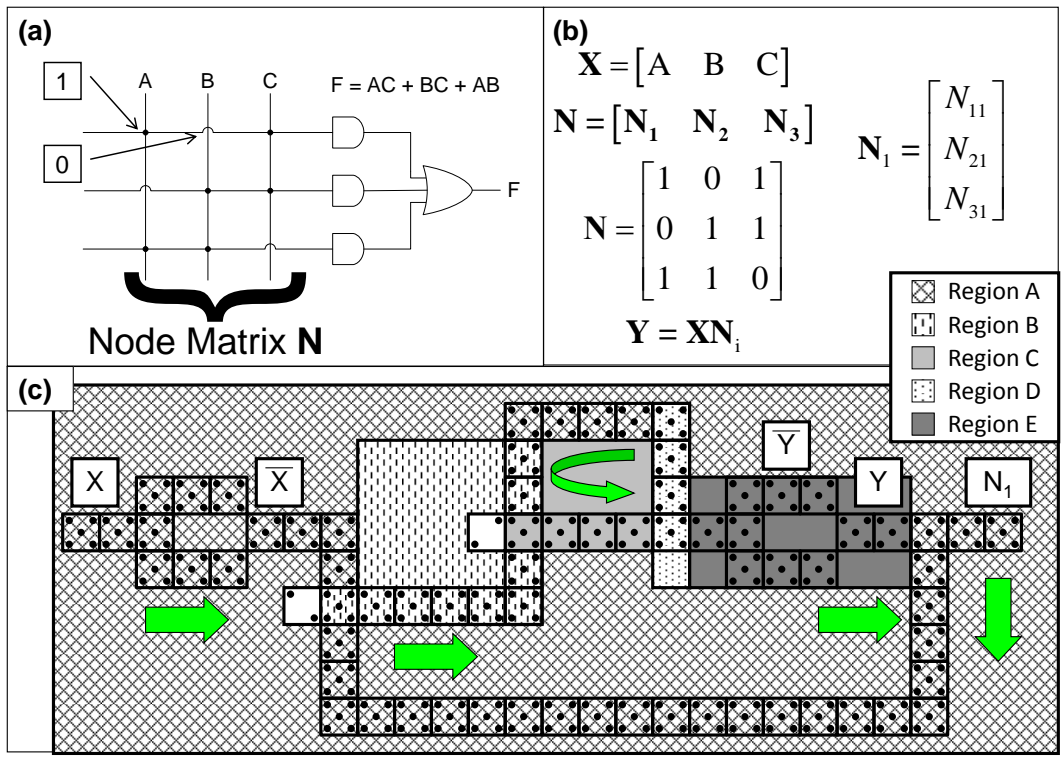

**Figure 9.** Schematic implementation of the programmable logic array (PLA) using a MM. (a) PLA for the digital circuit F = AC + BC + AB. The node matrix **N** is represented by ones at the nodes and zeroes elsewhere. Each node row fills a column vector in the node matrix, $N_i$. (b) Matrices for the PLA QCA matrix multiplier. The input digital values are given in the row vector **X**, and the node matrix **N** has node columns $N_1$, $N_2$, and $N_3$. The resulting function digital circuit F is given by the sum of the output matrices **Y**. (c) QCA matrix multiplier cells. The two inverters allow for the implementation of the PLA.

# Supplementary Information

# Matrix multiplication using quantum-dot cellular automata to implement conventional microelectronics

**Joshua D Wood[1,3] and P Douglas Tougaw[2]**

[1] Department of Electrical and Computer Engineering, Beckman Institute for Advanced Science and Technology, University of Illinois, Urbana-Champaign, IL 61801, USA
[2] Department of Electrical and Computer Engineering, Valparaiso University, Valparaiso, IN 46383, USA
[3] To whom correspondence should be addressed, e-mail: jwood2@illinois.edu

Sections:
**S. 1. Binary matrix multiplication**
**S. 2. Matrix multiplier state transition table**
**S. 3. Nine multiplier simulation result**
**S. 4. Proposed physical implementation of clocking regions**
**S. 5. Additional reset line for QCA matrix multiplier**

## S.1. Binary matrix multiplication

Consider the following 3x3 matrices **A** and **B**:

$$\mathbf{A} = \begin{bmatrix} 0 & 1 & 0 \\ 1 & 0 & 0 \\ 0 & 1 & 0 \end{bmatrix} \quad (S.1.1)$$

$$\mathbf{B} = \begin{bmatrix} 1 & 0 & 1 \\ 0 & 0 & 1 \\ 0 & 1 & 0 \end{bmatrix} \quad (S.1.2)$$

Binary matrix multiplication for the matrices **A** and **B** based on AND and OR gates is

$$\mathbf{AB} = \begin{bmatrix} 0*1+1*0+0*0 & 0*0+1*0+0*1 & 0*1+1*1+0*0 \\ 1*1+0*0+0*0 & 1*0+0*0+0*0 & 1*1+0*1+0*0 \\ 0*1+1*0+0*0 & 1*1+0*1+0*0 & 0*1+1*1+0*0 \end{bmatrix} = \begin{bmatrix} 0 & 0 & 1 \\ 1 & 0 & 1 \\ 0 & 1 & 1 \end{bmatrix} \quad (S.1.3)$$

which is the binary equivalent of conventional matrix multiplication in linear algebra. The result for each matrix element cannot vary from 0 or 1.

## S.2. Matrix multiplier state transition table

**Table S.2.1**. QCA matrix multiplier phase transitions for a three device QCA matrix multiplier. Note the "Block" state – this is a state where the potential barriers are very high to lower the probability of tunnelling.

| State | Phase | Region A | Region B | Region C | Region D | Region E |
|---|---|---|---|---|---|---|
|  | 1 | Locking | Relaxed | Relaxed | Relaxed | Relaxed |
|  | 2 | Locked | Relaxed | Relaxed | Relaxed | Locking |
| System reset | 3 | Relaxing | Relaxed | Relaxed | Locking | Locked |
|  | 4 | Locking | Relaxed | Relaxed | Locked | Block |
|  | 5 | Locked | Locking | Relaxed | Relaxing | Block |
|  | 6 | Relaxing | Locked | Locking | Relaxed | Block |
| 1st bit inputted | 7 | Relaxed | Relaxing | Locked | Locking | Block |
|  | 4 | Locking | Relaxed | Relaxing | Locked | Block |
|  | 5 | Locked | Locking | Relaxed | Relaxing | Block |
|  | 6 | Relaxing | Locked | Locking | Relaxed | Block |
| 2nd bit inputted | 7 | Relaxed | Relaxing | Locked | Locking | Block |
|  | 4 | Locking | Relaxed | Relaxing | Locked | Block |
|  | 5 | Locked | Locking | Relaxed | Relaxing | Block |
|  | 6 | Relaxing | Locked | Locking | Relaxed | Block |
| 3rd bit inputted | 7 | Relaxed | Relaxing | Locked | Locking | Block |
|  | 8 | Relaxed | Relaxed | Relaxing | Locked | Locking |
|  | 9 | Locking | Relaxed | Relaxed | Relaxing | Locked |
|  | 10 | Locked | Don't care | Don't care | Don't care | Don't care |

## S.3. Nine multiplier simulation result

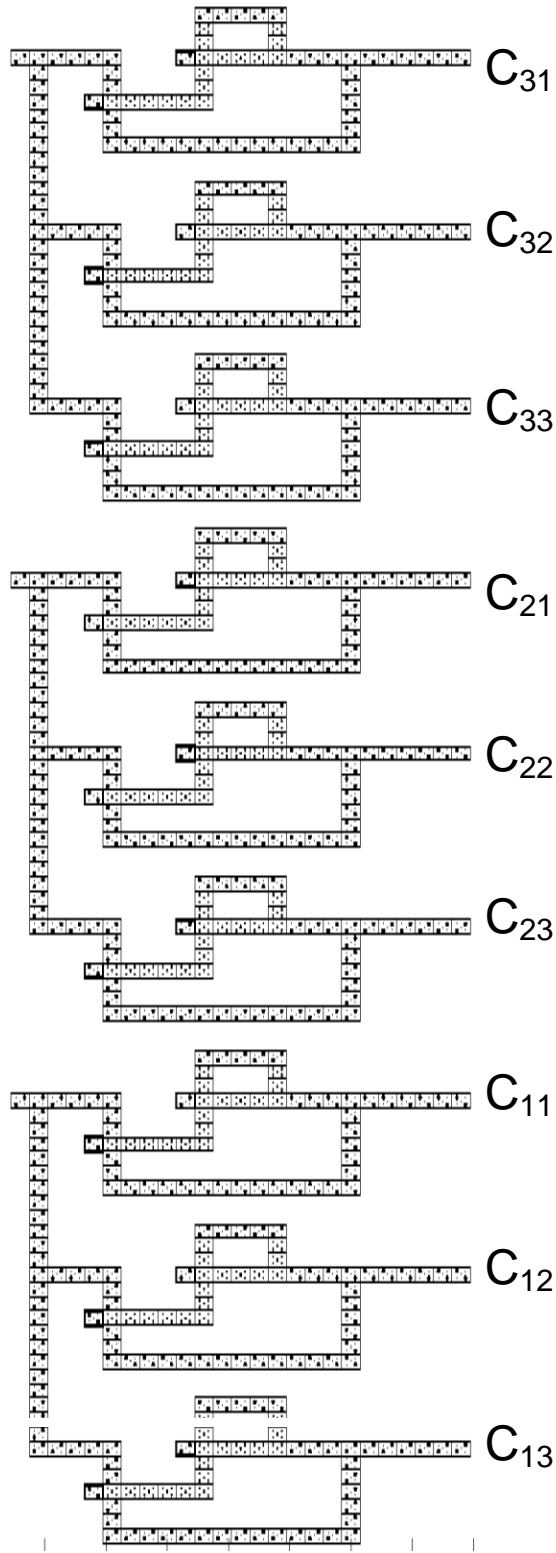

## S. 4. Proposed physical implementation of clocking regions

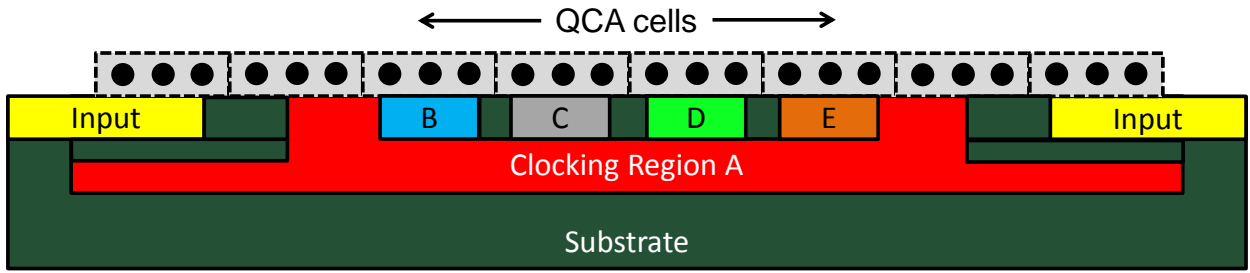

**Figure S-1.** Cross-section of QCA matrix multiplier clocking regions. In this schematic, metalized clocking regions will be layered and contacted out of the plane. Additionally, the QCA cells will be placed on top of these clocking regions, and inputs will pass into the cells by means of metal input contacts buried below the QDs. Multiple masks and evaporation steps could achieve this structure, provided one could control the placement of the top QCA cells.

## S. 5. Additional reset line for QCA matrix multiplier

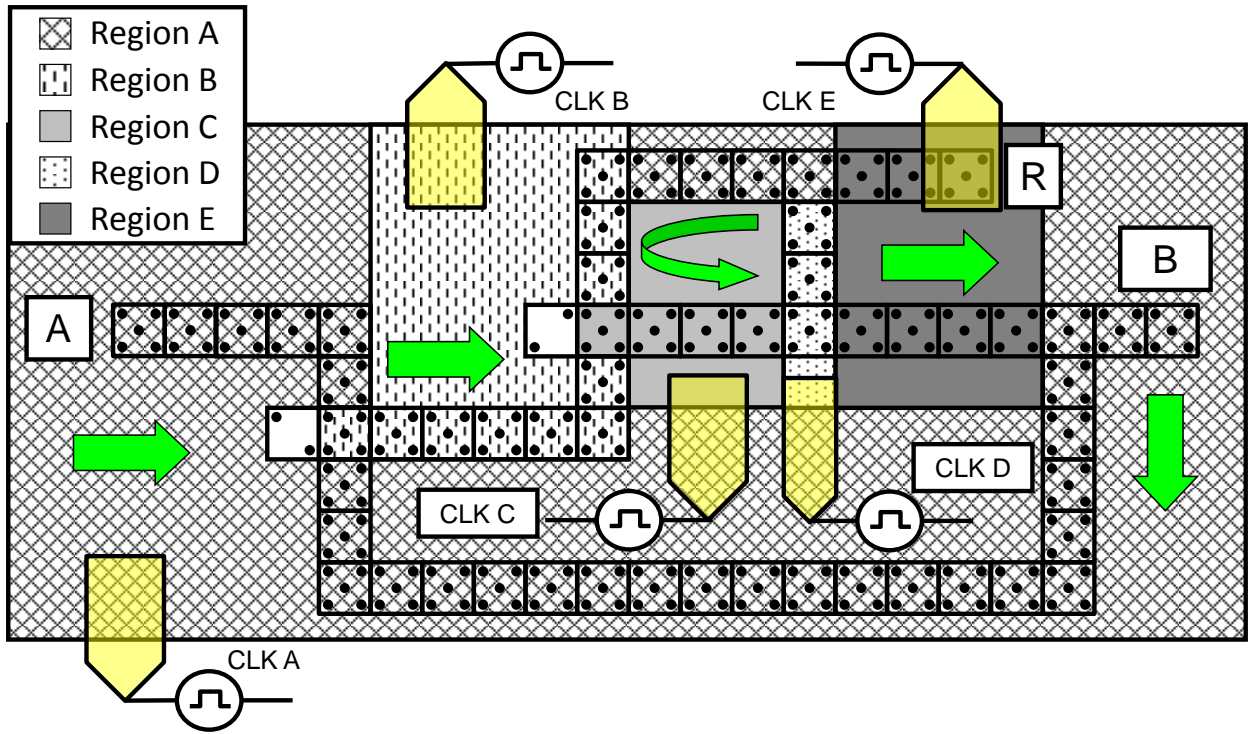

**Figure S-2.** Modified schematic of the QCA MM from Figure 4. This schematic shows possible locations for clocking contacts (e.g. CLK A). Additionally, the schematic shows an additional input line R, which allows for a reset of the OR loop in the MM. The addition of three QCA cells obviates two of the reset cycles required in the original conception of the QCA MM design, lowering the total cycle number to 15 for the one MM model. Nevertheless, adding cells complicates the design's compactness and becomes problematic with larger MM designs, like the nine multiplier formulation for 3 x 3 matrix multiplication.